\documentclass[preprint,12pt]{elsarticle}

\usepackage{graphicx}
\usepackage{amssymb}
\usepackage{amsmath}

\newcommand{\X}{\mathbf{X}}
\newcommand{\F}{\mathbf{F}}
\newcommand{\df}{\mathbf{df}}
\newcommand{\f}{\mathbf{f}}
\newcommand{\uvec}{\mathbf{u}}
\newcommand{\wvec}{\mathbf{w}}
\newcommand{\x}{\mathbf{x}}

\newcommand{\ep}{\varepsilon}

\newcommand{\norm}[1]{\lvert#1\rvert}

\begin{document}

\begin{frontmatter}



%%\title{Title\tnoteref{label1}}

\title{Aperiodic dynamics in a deterministic model of attitude formation in social groups}


\author{Jonathan A. Ward\corref{Corresponding Author}}
%% \address{Centre for the Mathematics of Human Behaviour, Department of
%%   Mathematics and Statistics, University of Reading, Whiteknights, UK}
\ead{j.a.ward@reading.ac.uk}

\author{Peter Grindrod}

\address{Centre for the Mathematics of Human Behaviour, Department of
  Mathematics and Statistics, University of Reading, Whiteknights, UK}

\begin{abstract}
Homophily and social influence are the fundamental mechanisms that
drive the evolution of attitudes, beliefs and behaviour within social
groups. Homophily relates the similarity between pairs of individuals'
attitudinal states to their frequency of interaction, and hence
structural tie strength, while social influence causes the convergence
of individuals' states during interaction.  Building on these basic
elements, we propose a new mathematical modelling framework to
describe the evolution of attitudes within a group of interacting
agents.  Specifically, our model describes sub-conscious attitudes
that have an activator-inhibitor relationship.  We consider a
homogeneous population using a deterministic, continuous-time
dynamical system.  Surprisingly, the combined effects of homophily and
social influence do not necessarily lead to group consensus or global
monoculture.  We observe that sub-group formation and polarisation-like
effects may be transient, the long-time dynamics being quasi-periodic
with sensitive dependence to initial conditions. This is due to the
interplay between the evolving interaction network and Turing
instability associated with the attitudinal state dynamics.
\end{abstract}


\end{frontmatter}

% \linenumbers


%%%%%%%%%%%%%%%%%%%%%%%%%%%%%%%%%%%%%%%%%%%%%%%%%%%%%%%%%%%%%%

% ONLY ADDITION!

\graphicspath{{../}}

%%%%%%%%%%%%%%%%%%%%%%%%%%%%%%%%%%%%%%%%%%%%%%%%%%%%%%%%%%%%%%

\section{Introduction}
\label{sec:introduction}
Our attitudes and opinions have a reciprocal relationship with those
around us: who we know depends on what we have in common, while
simultaneously our beliefs influence, and are influenced by, those of
our peers.  These two mechanisms---homophily and social
influence---underpin a wide range of social phenomena, including the
diffusion of innovations
\cite{Valente96,Watts02,Bettencourt06,Onnela10,Krapivsky11}, complex
contagions \cite{Dodds05,Centola07a,LopezPintado08}, collective action
\cite{Granovetter78,Fowler10,Baldassarri07}, opinion dynamics
\cite{Friedkin99,Hegselmann02,Ben-Naim03,Holme06,Gil06,Kozma08,Mobilia11,McCullen11,Durrett11}
and the emergence of social norms
\cite{Shoham97,Friedkin01,Centola05}. Thus homophily and social
influence represent the atomistic ingredients for models of social
dynamics \cite{Castellano09}.  Starting with these basic elements, we
investigate a new type of modelling framework intended to describe the
coevolution of sub-conscious attitudinal states and social tie
strengths in a population of interacting agents.

The first ingredient in our modelling framework, homophily, relates
the similarity of individuals to their frequency of interaction
\cite{McPherson01}.  Thus homophily is structural, affecting the
strength of ties between people and hence the underlying social
network.  Homophily has been observed over a broad range of
sociodemographics: implicit characteristics, such as age, gender and
race; acquired characteristics, such as education, religion and
occupation; and internal states that govern attitudes and behaviour
\cite{Lazarsfeld54,McPherson01}.  Homophily inextricably links state
dynamics with the evolution of social tie strength, and consequently a
faithful model must be coevolutionary, connecting both the dynamics
\emph{of} the social network and the dynamics \emph{on} the social
network \cite{Centola07}.  Such network models are known as
\emph{coevolving} or \emph{adaptive}; see \cite{Gross07} for a review.
There has been a recent surge of interest in coevolving networks,
particularly models of opinion dynamics
\cite{Holme06,Gil06,Kozma08,Durrett11}, which build on simple models
of voting behaviour.

The second ingredient is social influence, which affects people's
attitudinal state through typically dyadic interactions.  It is a
fundamental result of social psychology that people tend to modify
their behaviour and attitudes in response to the opinions of others
\cite{Sherif36,Campbell61,Collins70}, sometimes even when this
conflicts sharply with what they know to be true \cite{Asch51} or
believe to be morally justifiable \cite{Milgram65}.  Similarly to
Flache and Macy \cite{Flache11}, we use diffusion to model social
influence: agents adjust their state according to a weighted sum of
the differences between their state and their neighbours'.  The
weights, which represent the strength of influence between pairs of
agents, are the corresponding elements of the undirected (dynamic)
social network, whose evolution is driven by homophily.  Although our
model is built on the notions of homophily and social influence
described above, we point out that differentiating between the effects
of these processes, particularly in observational settings, may be
very difficult \cite{Aral09,Shalizi11}.

Social scientists have developed `agent-based' models that incorporate
homophily and social influence in order to examine a variety of
social-phenomena, including group stability \cite{Carley91}, social
differentiation \cite{Mark98} and cultural dissemination
\cite{Axelrod97}, where a culture is defined as an attribute that is
subject to social influence.  In such models, an agent's state is
typically described by a vector of discrete cultures and the more
similar (according to some metric) two agent's states are, the higher
the probability of dyadic interactions between them (homophily) in
which one agent replicates certain attributes of the other (social
influence).  Surprisingly, the feedback between homophily and social
influence does not necessarily lead to a global monoculture
\cite{Axelrod97}. In fact, the dissolution of ties between culturally
distinct groups, or equivalently the creation of `structural holes'
\cite{Macy03}, may lead to \emph{cultural polarisation}---equilibrium
states that preserve diversity.  However, such multi-cultural states
are not necessarily stable when there is `cultural drift', i.e. small,
random perturbations or noise, which inevitably drive the system
towards monoculture \cite{Klemm05}.  There have been a number of
attempts to develop models with polarised states that are stable in
the presence cultural drift \cite{Centola07,Flache11}, but this is
still an open area of research \cite{Castellano09}.

Two key features differentiate our approach from those described
above.  Firstly, we specifically focus on \emph{sub-conscious}
attitude formation driven by a general class of activator-inhibitor
processes.  This is motivated by neuropsychological evidence that the
activation of emotional responses are associated with the
(evolutionarily older) regions of the brain know as the limbic system
and our inhibitions are regulated by the (evolutionarily younger)
prefrontal cortex \cite{Morgane05}.  This has led psychologists to
develop theories in which various personality traits (such as
extraversion, impulsivity, neuroticism and anxiety) form an
independent set of dimensions along which different types of behaviour
may be excited or regulated \cite{Mathews99,Eysenck67,Gray87}.  Thus
it is natural in our modelling framework for these processes to be
communicated independently and in parallel through distinct
transmission channels and hence via distinct diffusion coefficients.
The consequence of activator-inhibitor attitudinal state dynamics is
that we would expect to encounter Turing instability, since the rates
at which social influence can change homophilious attributes may
differ dramatically.

Secondly, and in sharp contrast to recent models of cultural
dissemination \cite{Axelrod97,Centola07,Flache11} and indeed many
other types of behavioural model \cite{Castellano09} that are
stochastic or probabilistic\footnote{A notable exception is the
  deterministic, discrete-time model of Friedkin and Johnsen
  \cite{Friedkin99}; see also \cite{Hegselmann02} and references
  therein.}, we consider a \emph{deterministic}, continuous-time
dynamical systems formulation.  While this does not reflect the
mercurial nature seemingly ingrained in human interaction, it allows
us to probe the underlying mechanisms driving dynamical phenomena.  In
fact, our principle observation is that the tension between Turing
instability and the coevolution of the social network and attitudinal
states gives rise to aperiodic dynamics that are sensitive to initial
conditions and surprisingly unpredictable.  This begs the question,
are the mechanisms that govern our behaviour the cause of its
volatility?  For parsimony, we also consider systems of homogeneous
agents.  This allows us to identify parameters that destabilise the
global monocultural steady state, giving rise to transient sub-group
formation.

This paper is organised as follows: in Section~\ref{sec:model}, we
describe our model in detail, analyse the stability of global
monoculture and describe the underlying dynamical mechanisms; in
Section~\ref{sec:examples} we illustrate typical numerical results
from both a large population of individuals and a simple example
consisting of just two agents; in Section~\ref{sec:conclusion} we
summarise our work and finally in Section~\ref{sec:discussion} we
discuss our results in the context of other models of cultural
dynamics and polarisation phenomena.

\section{A deterministic model of cultural dynamics}
\label{sec:model}
Consider a population of $N$ identical individuals (agents/actors),
each described by a set of $M$ real attitude state variables that are
continuous functions of time $t$. Let $\x_i (t) \in \mathbb{R}^M $
denote the $i$th individual's attitudinal state. In the absence of any
influence or communication between agents we assume that each
individual's state obeys an autonomous rate equation of the form
\begin{equation}
\label{one}
\dot{\x}_{i}=\f(\x_i),\ \ i=1,...,N,
\end{equation}
where $\f$ is a given smooth field over $\mathbb{R}^M$, such that
$\f(\x^*)=0$ for some $\x^*$. Thus (\ref{one}) has a uniform
population equilibrium $\x_i=\x^*$, for $i=1,...,N$, which we shall
assume is locally asymptotically stable. As discussed in the
introduction, we shall more specifically assume that (\ref{one}) is
drawn from a class of activator-inhibitor systems.

Now suppose that the individuals are connected up by a dynamically
evolving weighted network. Let $A(t)$ denote the $N \times N$ weighted
adjacency matrix for this network at time $t$, with the $ij$th term,
$A_{ij}(t)$, representing the instantaneous weight (frequency and/or
tie strength) of the mutual influence between individual $i$ and
individual $j$ at time $t$. Throughout we assert that $A(t)$ is
symmetric, contains values bounded in [0,1] and has a zero diagonal
(no self influence). We extend (\ref{one}) and adopt a first order
model for the coupled system:
\begin{equation}
\label{two}
\dot{\x}_{i}=\f(\x_i)+ D \sum_{j=1}^NA_{ij}\left(\x_{j}-\x_{i}\right),\ \ \ i=1,...,N.
\end{equation}
Here $D$ is a real, diagonal and non-negative matrix containing the
maximal transmission coefficients (diffusion rates) for the
corresponding attitudinal variables between neighbours.  Thus some of
the attitude variables may be more easily or readily transmitted, and
are therefore influenced to a greater extent by (while simultaneously
being more influential to) those of neighbours.  Note that
$\x_i=\x^*$, for $i=1,...,N$, is also a uniform population equilibrium
of the augmented system.

Let $\X(t)$ denote the $M \times N$ matrix with $i$th column given by
$\x_i(t)$, and $\F(\X)$ be the $M \times N$ matrix with $i$th column
given by $\f(\x_i(t))$. Then (\ref{two}) may be written as
\begin{equation}
% \label{three}
\dot{\X}=\F(\X) - D \X \Delta.
\label{eq:Xmat}
\end{equation}
Here $\Delta(t)$ denotes the weighted graph Laplacian for $A(t)$,
given by $\Delta(t)={\rm diag}(\mathbf{k}(t))-A(t)$, where
$\mathbf{k}(t)\in\mathbb{R}^N$ is a vector containing the degrees of
the vertices ($k_{i}(t)=\sum_{j=1}^N
A_{ij}(t)$). Equation~(\ref{eq:Xmat}) has a rest point at $\X=\X^*$,
where the $i$th column of $\X^*$ is given by $\x^*$ for all
$i=1,...,N$.

To close the system, consider the evolution equation for $A(t)$ given
by
\begin{equation}
\dot{A}=\alpha A\circ({\bf 1}-A)\circ\left(\ep {\bf 1}-\Phi(\X)\right).
\label{eq:A}
\end{equation}
Here ${\bf 1}$ denotes the adjacency matrix of the fully weighted
connected graph (with all off-diagonal elements equal to one and all
diagonal elements equal to zero); $\circ$ denotes the element-wise
`Hadamard' matrix product; $\alpha>0$ is a rate parameter; $\ep>0$ is
a homophily scale parameter; and $\Phi:\mathbb{R}^{N\times
  N}\rightarrow\mathbb{R}^{N\times N}$ is a symmetric matrix function
that incorporates homophily effects. We assume $\Phi$ to be of the
form $\Phi_{ij}:=\phi(\norm{\x_{i}-\x_{j}})\ge0$, where $\norm{\cdot}$
is an appropriate norm or semi-norm, and the real function $\phi$ is
monotonically increasing with $\phi(0)=0$.  Note that the sign of the
differences held in $\ep{\bf 1}-\Phi(X)$ controls the growth or decay
of the corresponding coupling strengths. All matrices in (\ref{eq:A})
are symmetric, so in practice we need only calculate the
super-diagonal terms.  For the $ij$th edge, from (\ref{eq:A}), we have
\begin{equation*}
\dot{A}_{ij}=\alpha{A}_{ij}(1-{A}_{ij})(\ep-\phi(\norm{\x_i-\x_j}).
\end{equation*}
The nonlinear ``logistic growth"-like term implies that the weights
remain in [0,1], while we refer to the term
$\ep-\phi(\norm{\x_i-\x_j})$ as the {\it switch} term.

\subsection{Stability analysis}
\label{sec:stability}
By construction, there are equilibria at $\X=\X^*$ with either $A=0$ or
$A={\bf 1}$.  To understand their stability, let us assume that
$\alpha\to 0$ so that $A(t)$ evolves very slowly. We may then consider
the stability of the uniform population, $\X^*$, under the fast dynamic
(\ref{eq:Xmat}) for any fixed network $A$.  Assuming that $A$ is constant,
writing $\X(t)=\X^* + \tilde{\X}(t)$ and Linearising (\ref{eq:Xmat})
about $\X^*$, we obtain
\begin{equation}
\label{thic}
\dot{\tilde{\X}}=\df(\x^*) \tilde{\X}- D \tilde{ \X} \Delta.
\end{equation}
Here $\df(\x^*)$ is an $M \times M$ matrix given by the linearisation
of $\f(\x)$ at $\x^*$.  Letting $(\lambda_i, \wvec_i)\in [0,\infty)
  \times \mathbb{R}^N,\ i=1,...,N$, be the eigen-pairs of $\Delta$,
  then we may decompose uniquely \cite{nakao10}:
$$\tilde{\X}(t) =\sum_{i=1}^N \uvec_i(t) \wvec_i^T,$$ where each
  $\uvec_i(t)\in \mathbb{R}^M$.  The stability analysis of
  (\ref{thic}) is now trivial since decomposition yields
$$\dot {\uvec}_i=( \df(\x^*)- D \lambda_i ) \uvec_i.$$ Thus the
  uniform equilibrium, $\X^*$, is asymptotically stable if and only if
  all $N$ matrices, $( \df(\x^*)- D \lambda_i )$, are simultaneously
  stability matrices; and conversely is unstable in the $i$th mode of
  the graph Laplacian if $( \df(\x^*)- D \lambda_i )$ has an
  eigenvalue with positive real part.

Consider the spectrum of $( \df(\x^*)- D \lambda)$ as a function of
$\lambda$. If $\lambda$ is small then this is dominated by the
stability of the autonomous system, $ \df(\x^*)$, which we assumed to
be stable. If $\lambda$ is large then this is again a stability
matrix, since $D$ is positive definite. The situation, dependent on
some collusion between choices of $D$ and $\df(\x^*)$, where there is
a {\em window of instability} for an intermediate range of $\lambda$,
is know as a Turing instability. Turing instabilities occur in a
number of mathematical applications and are tied to the use of
activator-inhibitor systems (in the state space equations, such as
(\ref{one}) here), where inhibitions diffuse faster than activational
variables. 

Now we can see the possible tension between homophily and Turing
instability in the attitude dynamics when the timescale of the
evolving network, $\alpha$, is comparable to the changes in agents'
states.  There are two distinct types of dynamical behaviour at work.
In one case, $\Delta(t)$ has presently no eigenvalues within the
window of instability and each individual's states $\x_i(t)$ approach
the mutual equilibrium, $\x^*$; consequently all switch terms become
positive and the edge weights all grow towards unity, i.e. the fully
weighted clique.  In the alternative case, unstable eigen-modes cause
the individual states to diverge from $\x^*$, and subsequently some of
the corresponding switch terms become negative, causing those edges to
begin losing weight and hence partitioning the network.

The eigenvalues of the Laplacian for the fully weighed clique, $A={\bf
  1}$, are at zero (simple) and at $N$ (with multiplicity $N-1$).  So
the interesting case is where the system parameters are such that
$\lambda=N$ lies within the window of instability. Then the steady
state $(\X, A)=(\X^*, {\bf 1})$ is unstable and thus state variable
patterns will form, echoing the structure of (one or many of) the
corresponding eigen-mode(s).  This Turing driven symmetry loss may be
exacerbated by the switch terms (depending upon the choice of $\ep$
being small enough), and then each sub-network will remain relatively
well intra-connected, while becoming less well connected to the other
sub-networks.  Once relatively isolated, individuals within each of
these sub-networks may evolve back towards the global equilibrium at
$\x^*$, providing that $A(t)$ is such that the eigenvalues of $\Delta$
have by that time left the window of instability. Within such a less
weightily connected network, all states will approach $\x^*$, the
switch terms will become positive, and then the whole qualitative
cycle can begin again.

Thus we expect aperiodic or pseudo-cyclic emergence and diminution of
patterns, representing transient variations in attitudes in the form
of different {\em norms} adopted by distinct sub-populations. As we
shall see though, the trajectory of any individual may be sensitive
and therefore effectively unpredictable, while the dynamics of the
global behaviour is qualitatively predictable.

In the next section we introduce a specific case of the more general
setting described here.

\section{Examples}
\label{sec:examples}
We wish to consider activator-inhibitor systems as candidates for the
attitudinal dynamics in (\ref{one}) and hence (\ref{eq:Xmat}). The
simplest such system has $M=2$, with a single inhibitory variable,
$x(t)$, and a single activational variable, $y(t)$. Let
$\x_i(t)=(x_i(t), y_i(t))^T$ in (\ref{two}), and consider the
Scnackenberg dynamics defined by the field
\begin{equation}
\f(\x_i)=(p-x_i y_i^2,\, q-y_i+x_iy_i^2)^T,
\end{equation}
where $p >  q\ge 0$ are constants. The equations have
the required equilibrium point at
\begin{equation}
\x^*=\left(\frac{p}{(p+q)^2},\, p+q\right)^T,
\end{equation}
and in order that $\df $ be a stability matrix, we must have
$$ p-q<(p+q)^3.$$ We employ $\phi_{ij}=(x_i-x_j)^2$ as the homophily
function and we must have $D={\rm diag}(D_1, D_2)$ in (\ref{eq:Xmat}).

When $M=2$, the presence of Turing instability depends on the sign of
the determinant of $( \df(\x^*)-D \lambda)$, which is quadratic in
$\lambda$. For the Schnakenberg dynamics defined above, the roots of
this quadratic are given by
\begin{equation*}
\lambda_{\pm}= \frac{  (p-q) - \frac{D_2}{D_1} (p+q)^3  \pm \sqrt{    \left[(p-q)-\frac{D_2}{D_1} (p+q)^3  \right]^2  -  4\frac{D_2}{D_1}  (p+q)^4}
  }{2 D_2  (p+q)}   
   >0. 
\end{equation*}
It is straightforward to show that if 
\begin{equation}
\frac{D_2}{D_1}<\frac{3p+q-2\sqrt{2p(p+q)}}{(p+q)^3}:=\sigma_{\rm c},
\end{equation}
then $\lambda_{\pm}$ are real positive roots and hence $( \df(\x^*)-
D \lambda)$ is a stability matrix if and only if $\lambda$ lies
outside of the interval $(\lambda_-, \lambda_+)$, the {\em window of
  instability}. Inside there is always one positive and one negative
eigenvalue, and the equilibrium $\X^*$ is unstable for any fixed
network $A$.  Note that, as is well known, it is the \emph{ratio} of
the diffusion constants that determines whether there is a window of
instability.

\subsection{Group dynamics}
\label{sec:group}
We now present simulations of the Schnakenberg dynamics with $N=10$.
Parameter values are $p=1.25$, $q=0.1$, $\alpha=10^{4}$,
$\varepsilon=10^{-6}$, $D_1\approx0.571$ and $D_2\approx0.037$. The
ratio of the diffusion constants is $D_2/D_1:=\sigma=0.9\sigma_{\rm
  c}$, and to ensure that the window of instability is centred on the
fully coupled system we have
\begin{equation}
D_1=\frac{(p-q)-\sigma(p+q)^3 }{2\sigma N(p+q)}.
\label{eq:centre}
\end{equation}
The initial coupling strengths were chosen uniformly at random between
0.1 and 0.5.  The initial values of $x$ and $y$ were chosen at equally
spaced intervals on a circle of radius $10^{-3}$ centred on the
uniform equilibrium. 

\begin{figure}[t]
\centering
\includegraphics[width=0.55\textwidth]{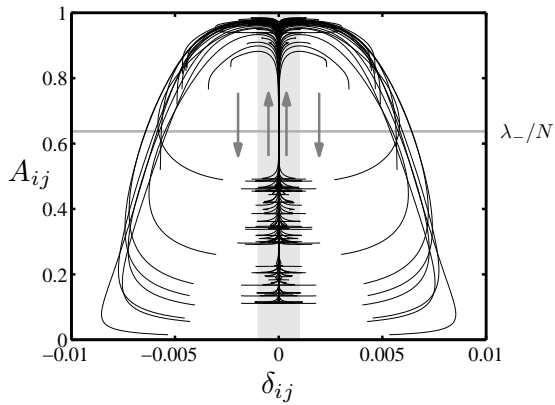}
\caption{Trajectories of $\delta_{ij}:=x_i-x_j$ and $A_{ij}$ for all
  $(i,j)$ pairs for unstable parameters integrated until
  $t\approx440$.  Parameter values and initial conditions are
  described in the main text. In the light grey shaded region,
  $\delta_{ij}<\varepsilon$ and the direction of trajectories are
  indicated with arrows.  The grey horizontal line indicates the
  scaled stability threshold (unstable above, stable below).
\label{fig:duA}}
\end{figure}

In Figure~\ref{fig:duA} we illustrate the trajectories of
$\delta_{ij}:=x_i-x_j$ and the corresponding coupling strengths
$A_{ij}$ up to $t\approx440$.
The shaded region corresponds to $\delta_{ij}<\varepsilon$, within
which the $A_{ij}$ increase and outside of which they decrease,
indicated by the dark grey arrows.  The horizontal grey line marks the
scaled instability threshold $\lambda_-/N$, which is indicative of the
boundary of instability, above being unstable and below being stable.
Because agents are only weakly coupled initially, their attitudes move
towards the steady state $\x^*$, which causes the differences
$\delta_{ij}$ to decrease. The switch terms subsequently become
positive and hence the coupling strengths increase, along with the
eigenvalues of the Laplacian $\lambda_i$.  When one or more of the
$\lambda_i$ are within the window of instability, some of the
differences $\delta_{ij}$ begin to diverge.  However, this eventually
causes their switch terms to become negative, reducing the
corresponding coupling strengths and hence some of the $\lambda_i$.
This then affects the differences $\delta_{ij}$, which start to
decrease, completing the qualitative cycle.  As the system evolves
beyond $t>440$, this quasi-cyclic behaviour becomes increasingly erratic.

\begin{figure}[p]
\centering
\includegraphics[width=0.79\textwidth]{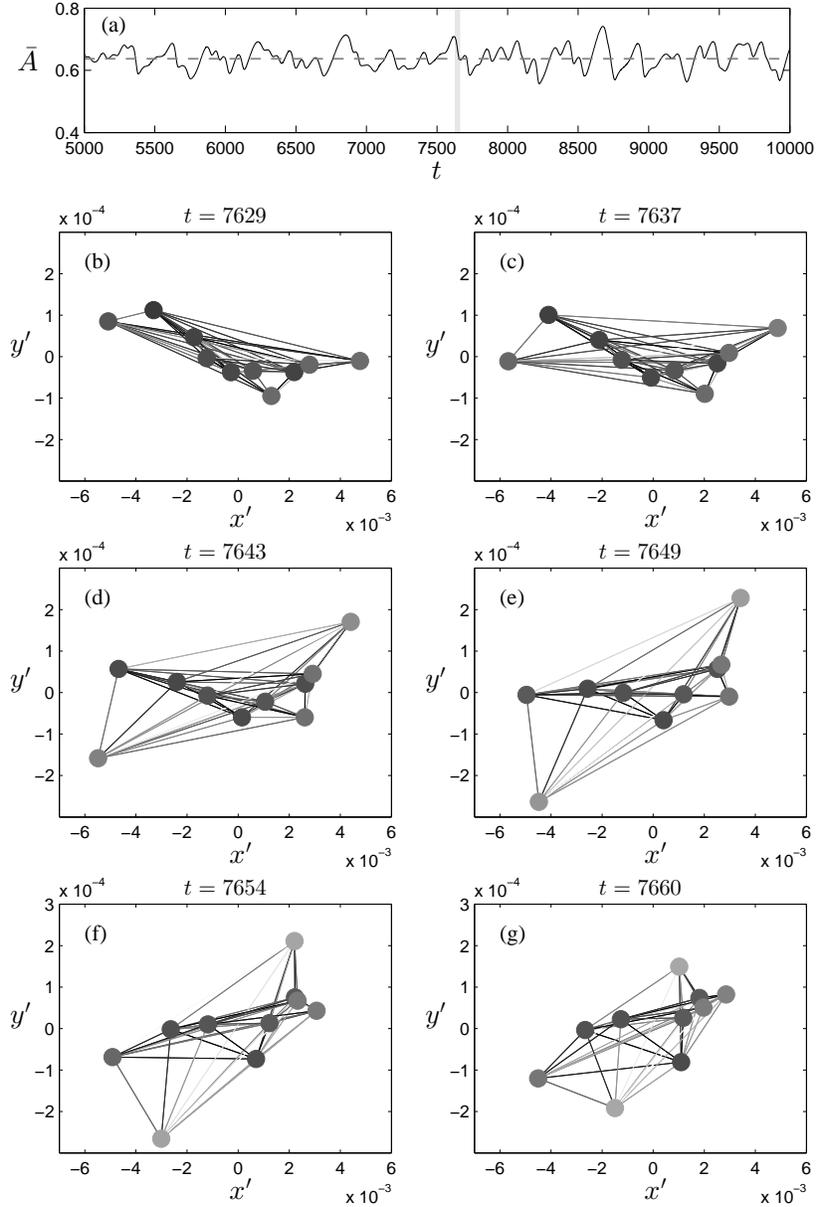}
\caption{Panel (a): mean coupling strength, $\bar{A}(t)$, time
  series. Panels (b)--(g): network snapshots at sequential time
  intervals. Node positions are plotted in the rotated coordinates
  $(x^{\prime},y^{\prime})$, shading illustrates coupling strength for
  edges and mean coupling strength for nodes.
\label{fig:network}}
\end{figure}

Although the long term behaviour of any given agent is unpredictable,
the behaviour of the mean coupling strength of the system fluctuates
around the instability boundary $k_{-}/N$. In
Figure~\ref{fig:network}(a), we plot the time series of the mean
coupling strength, $\bar{A}(t)$, between $t=5\times 10^3$ and
$10^{4}$.
The dashed line indicates the instability boundary
$k_{-}/N\approx0.6372$, which is very close to the time-averaged mean
coupling strength $\langle A \rangle\approx0.6343$.  Also plotted in
Figure~\ref{fig:network} are snapshots of the network at six
sequential times.  To improve the visualisation of the network, the
positions of nodes have been rotated by approximately $72^{\circ}$,
since the differences in diffusion rates mean that the unrotated
coordinates, $(x,y)$, become contracted in one direction.  The shading
of the nodes corresponds to their average coupling strength and the
shading of the edges correspond to their weight. The sequence of
figures illustrate the general scenario: agents' trajectories cycle
around the origin with the network repeatedly contracting and
expanding as agents become more and less similar in attitude
respectively.

\begin{figure}
\centering
\includegraphics[width=0.55\textwidth]{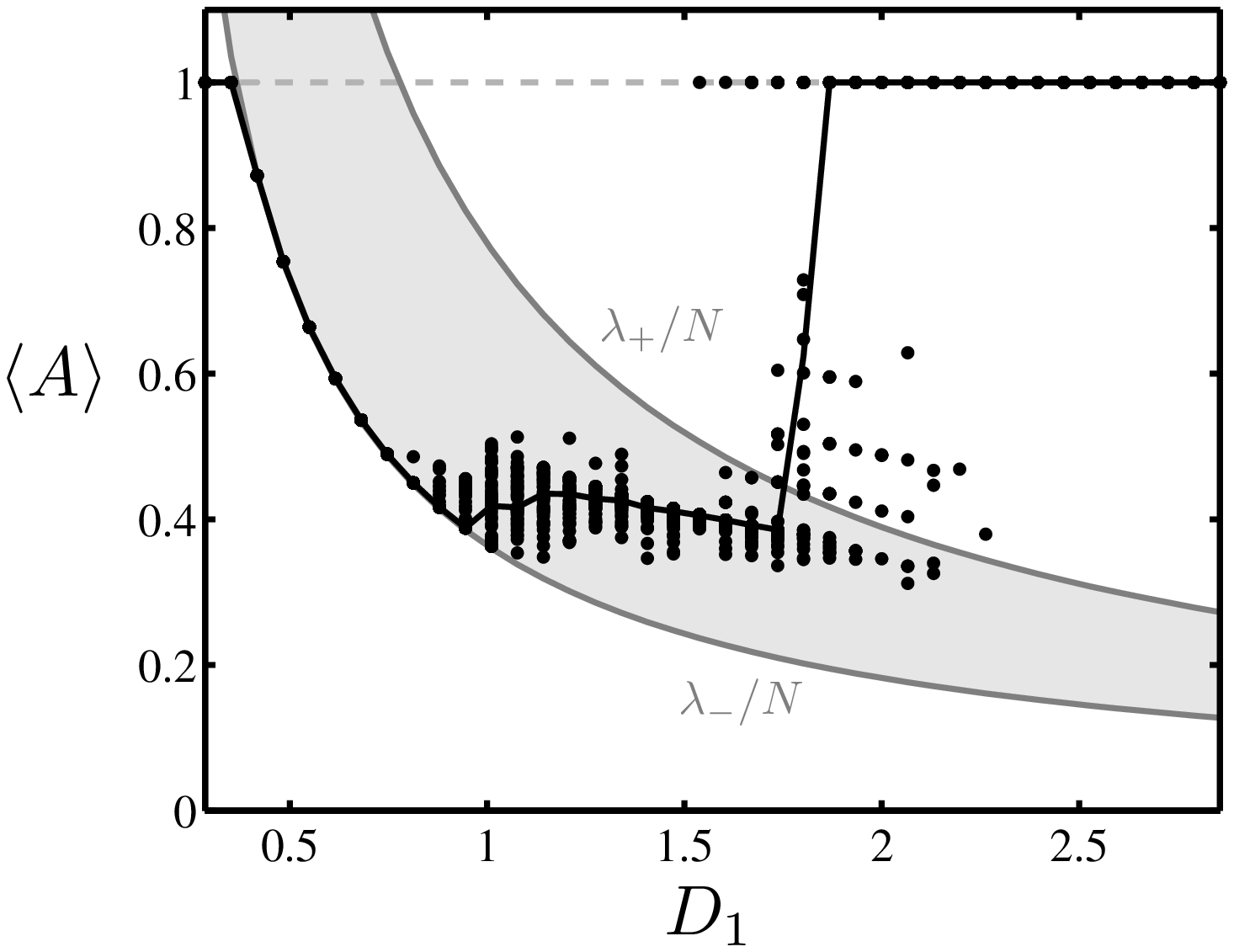}
\caption{Mean coupling strength at large times for different values of
  $D_1$. Shading indicates the window of instability, the dashed grey
  line indicates where $A={\bf 1}$. The black line is the median of 50
  realisations and the black markers are the mean coupling strengths
  for each of the realisations.
\label{fig:bif}}
\end{figure}

We now illustrate how the quasi-equilibrium end state changes as the
window of instability is moved. We fix all parameters as above, but
consider a range of values of $D_1$ whilst keeping the ratio
$D_2/D_1=0.9\sigma_{\rm c}$ held fixed. This has the effect of
shifting the window of instability from above $\langle A\rangle=1$ to
below as $D_1$ increases. We integrate until $t=1.5\times10^{5}$ and
then compute the mean coupling strength $\langle A\rangle$ for
$t\ge10^{4}$. We compute 50 realisations for each set of parameters,
the results of which are plotted in Figure~\ref{fig:bif}.
The shaded region corresponds to the (scaled) region of instability,
the dashed grey line is where $\langle A\rangle=1$, the black line is
the median of the fifty realisations and the dots are the values from
each of the realisations. At low values of $D_1$, where the $A={\bf
  1}$ equilibrium first becomes unstable, the mean coupling strength
fits tightly to the lower edge of the instability boundary at
$\lambda_{-}/N$. When the $A={\bf 1}$ equilibrium restabilises (at
$D_1\approx0.78$), the long time behaviour of the mean coupling
strength changes, moving away from the $\lambda_{-}/N$ boundary.  In
the region between $D_1\approx 1.5$ and $D_1\approx2.3$, some
realisations return to the fully coupled equilibrium $A={\bf 1}$, but
not all.  We would expect that simulating for longer would result in
more realisations reaching the fully coupled equilibrium, although it
is possible that its basin of attraction does not include every
initial condition in the set that we are sampling from.

\subsection{Dyad dynamics}
\label{sec:dyad}
To probe the mechanism driving the aperiodic dynamics illustrated in
Section~\ref{sec:group}, we consider a simpler dynamical setting
consisting of just two agents.  This reduces the coupling strength
evolution (\ref{eq:A}) to a single equation, and hence five equations
in total,
\begin{align}
\dot{x}_1&=p-x_1y_1^2-D_1a(x_1-x_2),\label{eq:dyadx1}\\
\dot{x}_2&=p-x_2y_2^2+D_1a(x_1-x_2),\\
\dot{y}_1&=q-y_1+x_1y_1^2-D_2a(y_1-y_2),\\
\dot{y}_2&=q-y_2+x_2y_2^2+D_2a(y_1-y_2),\label{eq:dyady2}\\
\dot{a}&=\alpha a(1-a)\left[\varepsilon-(x_1-x_2)^2\right].\label{eq:dyada}
\end{align}

\begin{figure}[p]
\centering
\includegraphics[width=0.99\textwidth]{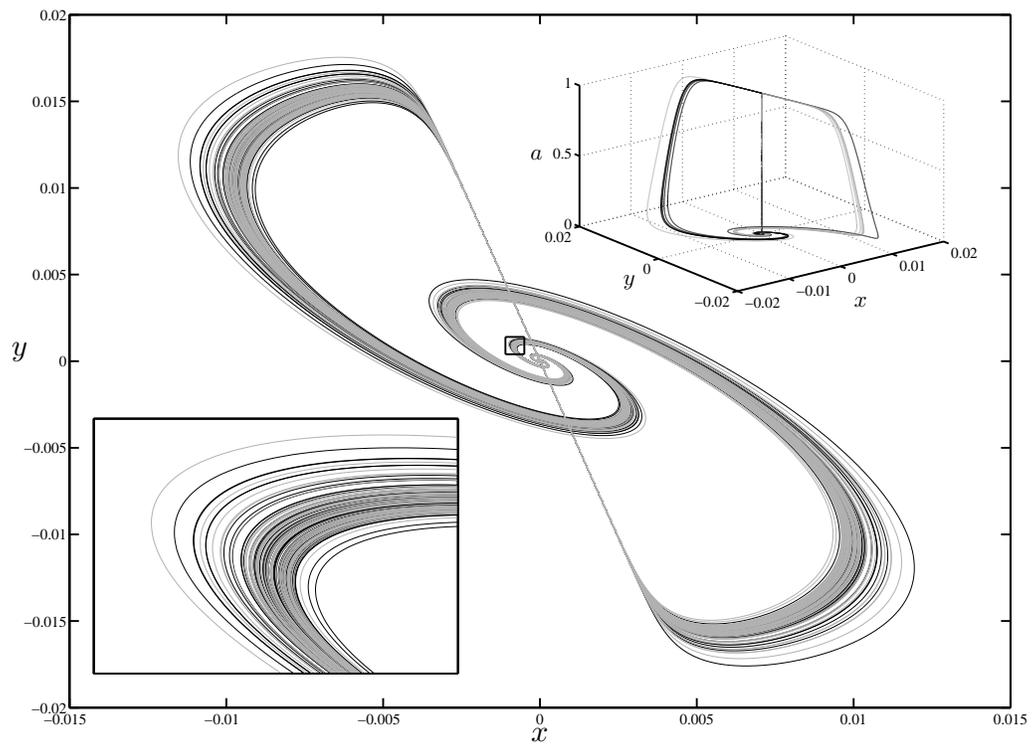}
\caption{Main: Trajectories of dyadic system in $(x,y)$ space with
  unstable parameters.  Upper-right inset: Trajectories in $(x,y,a)$
  space.  Lower-left inset: zoom of boxed region in the main plot.
  Parameters described in the main text.
\label{fig:chaos}}
\end{figure}

In Figure~\ref{fig:chaos}, we plot the trajectories for each of the
two agents (black and grey lines) in $(x,y)$ space, and in $(x,y,a)$
space in the upper-right inset.
The parameter values are $p=1.25$, $q=0.1$, $\alpha=10^{4}$,
$\varepsilon=10^{-6}$, $D_1\approx2.857$ and $D_2\approx0.184$. Again,
the diffusion constants have the ratio $D_2/D_1=0.9\sigma_{\rm c}$ and
the window of instability is centred on the fully coupled system via
(\ref{eq:centre}).  The initial conditions are chosen near to the
uniform equilibrium $\mathbf{x}^*=(x^*,y^*)^T$, specifically
$x_1(0)=x^*+1.5\times10^{-4}$, $x_2(0)=x^*-1.5\times10^{-4}$,
$y_1(0)=y^*-1\times10^{-6}$, $y_2(0)=y^*+1\times10^{-6}$; the initial
coupling strength is $a(0)=0.1$.  This system is numerically
stiff---on each cycle, trajectories get very close to the equilibrium
$\x^*$ and the invariant planes $a=0$ and $a=1$---thus very low error
tolerances are necessary in order to accurately resolve the
trajectories.

The mechanisms driving the near cyclic behaviour illustrated in
Figure~\ref{fig:chaos} are qualitatively similar to those described in
Section~\ref{sec:stability}, but the present case is much simpler
since the coupling constant, $a$, is a scalar.  If we consider $a$ as
a parameter in the attitudinal dynamics
(\ref{eq:dyadx1}--\ref{eq:dyady2}), then a Turing instability occurs
as a pitchfork bifurcation at some $a=a_*$, where $0<a_*<1$. The
equilibria at $(\x,a)=(\x^*,0)$ and $(\x^*,1)$ are both saddle-foci,
where the unstable manifolds are respectively parallel to the $a$-axis
and entirely within the attitudinal state space, $\x$.  Near to the
$(\x^*,1)$ equilibrium, a given trajectory tracks the unstable
manifold of $(\x^*,1)$ in one of two opposing directions, the choice
of which is sensitively dependent on its earlier position when
$a\approx a_*$.  The combination of this sensitivity together with the
spiral dynamics around the unstable manifold of $(\x^*,0)$, leads to
an orbit switching sides unpredictably on each near-pass of $(\x^*,1)$
(c.f. the top-right inset of Figure~\ref{fig:chaos}).  The mechanism
by which this chaotic behaviour arises is not standard (e.g. via a
Shilnikov bifurcation or homoclinic explosion) and warrants its own
study, which we address in an article currently in preparation.

\section{Conclusions}
\label{sec:conclusion}
We have proposed a new modelling framework to describe the evolution
of sub-conscious attitudes within social groups. We based this
framework on the fundamental mechanisms of homophily and social
influence, but it differs from previous approaches in two
respects. Firstly, we have focused on sub-conscious attitudes, where
it is natural to consider dynamics described by a class of
activator-inhibitor processes.  Secondly, we have formulated a
deterministic system, enabling us to highlight (via mathematical
analysis and simulation) the mechanisms driving dynamical phenomena.
Specifically, we have illustrated that the tension between Turing
instability and the evolving network topology gives rise to behaviour
that at the system level is qualitatively predictable --- sub-group
formation and dissolution --- yet at the level of individual agent
journeys is entirely unpredictable.  We point out that a stochastic
model based on similar principles to those described in this paper is
presented in \cite{Parsons12}, where qualitatively similar dynamical
phenomena are also observed.  Thus we might conclude that even if
stochasticity is entirely absent, the mechanisms that govern human
behaviour seem to give rise to unpredictable dynamics.

\section{Discussion}
\label{sec:discussion}
While we have differentiated our modelling framework from other models
of attitudinal dynamics \cite{Axelrod97,Centola07,Flache11}, we now
discuss our findings in the context of more general cultural models
and cultural polarisation.

Current interest in cultural models largely stems from the work of
Axelrod \cite{Axelrod97}, who demonstrated that local convergence
could lead to cultural polarisation.  This topic has particular
resonance in our digital society: will global connectivity accelerate
a descent into monoculture, or can diversity persist?  Models such as
Axelrod's provide us with an optimistic outlook, suggesting that even
the most basic mechanisms that model social influence and homophily
can lead to cultural diversity.  But by no means is there presently a
completely satisfactory understanding of this phenomena.  The
polarised states of the Axelrod model are fragile; even low rates of
random perturbations to cultural traits can reinstate global
monoculture \cite{Klemm05}.  Thus additional dynamical rules have been
investigated in this context.  The variant of the Axelrod model
proposed by Centola et al. \cite{Centola07} allows agents to
disassociate themselves from neighbours that have no similar traits
and select a new neighbour at random.  Similarly, a number of adaptive
network models of opinion dynamics have also found absorbing polarised
states, in which groups with differing opinions are completely
disconnected \cite{Gil06,Kozma08,Durrett11}.  Such polarised states
seem artificial and we conjecture that a form of cultural drift,
characterised by random rewiring of a small number of edges, would
destabilise these states.

This touches on another key issue: in the absence of noise, the
mechanisms employed by cultural dissemination models typically
\emph{reduce} diversity.  It is not surprising then that these types
of models can be perturbed in such a way that the eventual result is
monoculture.  An approach that allows diversity to increase has been
suggested by Flache and Macy \cite{Flache11}. They model social
influence via diffusion, whereby an agent adjusts their cultural
state, described by a vector of continuous real variables, according
to a weighted sum of the differences between their state and their
neighbours'.  The weights are dynamic and their evolution is driven by
homophily.  In some sense, the corresponding elements of our model are
like a continuous-time version of the Flache and Macy model. However,
Flache and Macy consider the weights embedded on a clustered network
and, more importantly, allow their weights to be negative,
representing \emph{xenophobia}.  It is this feature that allows
diversity to both decrease and increase via diffusion and convergence
respectively.  The effects of cultural drift on polarised states in
the Flache and Macy model have not been investigated in detail, but
perturbations can cause agents to switch groups \cite{Flache11} and so
we would expect that sustained noise would erode smaller groups.
Monoculture is also a stable fixed point of their model.

The key element that differentiates our model is that agents' cultural
states have an activator-inhibitor dynamic that is independent of
other agents.  The presence of diffusion allows for Turing instability
and hence means that diversity can increase. Moreover, we can identify
regions in which global monoculture is \emph{unstable}.  For fixed or
slowly evolving networks, instability gives rise to stable `Turing
patterns' \cite{nakao10}, which could be interpreted as culturally
polarised states. However, one would expect inter-group connections to
be weaker than intra-group connections within polarised states. But if
homophily dissolves such inter-group ties then the patterned or
polarised states can no longer be stable, since it is precisely the
differences in culture that balance individuals' attitudinal dynamics
with diffusion.  If non-trivial stable equilibria were to exist in our
model, they would involve a delicate balance of cultural differences
within the switching terms. However, we have seen no evidence of this
occurring in numerical simulations.  Thus in its present form,
sub-group formation and polarisation are transient phenomena in our
model.

It is possible however that extensions to our model could produce
stable polarised states. For example, introducing agent heterogeneity,
in the form of distinct uncoupled equilibria, offers some promise.
Agents could then adopt a state close to their uncoupled equilibrium,
allowing distinct groups to form, but Turing instability could still
destabilise the monocultural equilibrium.  Alternatively, the network
evolution equations could include higher order effects such as edge
snapping \cite{DeLellis10} or triangulation. These ideas will be
investigated further in follow-up work and we hope that our model may
provide a new paradigm from which to explore cultural polarisation
phenomena.

\section*{Acknowledgements}
JAW and PG acknowledge the EPSRC for support through MOLTEN
(EP/I016058/1). We would like to thank Jon Dawes, Michael Macy and
those at the Centre for Mathematics of Human Behaviour for their
valuable input, discussion and comments.

%%%%%%%%%%%%%%%%%%%%%%%%%%%%%%%%%%%%%%%%%%%%%%%%%%%%%%%%%%%%%%



%% References


%% \bibliographystyle{model1-num-names}
%% \bibliography{norms}

\begin{thebibliography}{48}
\expandafter\ifx\csname natexlab\endcsname\relax\def\natexlab#1{#1}\fi
\providecommand{\bibinfo}[2]{#2}
\ifx\xfnm\relax \def\xfnm[#1]{\unskip,\space#1}\fi
%Type = Article
\bibitem[{Valente(1996)}]{Valente96}
\bibinfo{author}{T.~W. Valente},
\newblock \bibinfo{title}{Social network thresholds in the diffusion of
  innovations},
\newblock \bibinfo{journal}{Soc. Networks} \bibinfo{volume}{18}
  (\bibinfo{year}{1996}) \bibinfo{pages}{69--89}.
%Type = Article
\bibitem[{Watts(2002)}]{Watts02}
\bibinfo{author}{D.~J. Watts},
\newblock \bibinfo{title}{A simple model for global cascades on random
  networks},
\newblock \bibinfo{journal}{Proc. Natl. Acad. Sci. U.S.A.} \bibinfo{volume}{99}
  (\bibinfo{year}{2002}) \bibinfo{pages}{5766--5771}.
%Type = Article
\bibitem[{Bettencourt et~al.(2006)Bettencourt, Cintron-Arias, Kaiser, and
  Castillo-Chavez}]{Bettencourt06}
\bibinfo{author}{L.~M.~A. Bettencourt}, \bibinfo{author}{A.~Cintron-Arias},
  \bibinfo{author}{D.~I. Kaiser}, \bibinfo{author}{C.~Castillo-Chavez},
\newblock \bibinfo{title}{The power of a good idea: Quantitative modeling of
  the spread of ideas from epidemiological models},
\newblock \bibinfo{journal}{Physica A} \bibinfo{volume}{364}
  (\bibinfo{year}{2006}) \bibinfo{pages}{513--536}.
%Type = Article
\bibitem[{Onnela and Reed-Tsochas(2010)}]{Onnela10}
\bibinfo{author}{J.~Onnela}, \bibinfo{author}{F.~Reed-Tsochas},
\newblock \bibinfo{title}{Spontaneous emergence of social influence in online
  systems},
\newblock \bibinfo{journal}{Proc. Natl. Acad. Sci. U.S.A.}
  \bibinfo{volume}{107} (\bibinfo{year}{2010}) \bibinfo{pages}{18375--18380}.
%Type = Article
\bibitem[{Krapivsky et~al.(2011)Krapivsky, Redner, and Volovik}]{Krapivsky11}
\bibinfo{author}{P.~L. Krapivsky}, \bibinfo{author}{S.~Redner},
  \bibinfo{author}{D.~Volovik},
\newblock \bibinfo{title}{Reinforcement-driven spread of innovations and fads},
\newblock \bibinfo{journal}{Journal of Statistical Mechanics: Theory and
  Experiment} \bibinfo{volume}{12} (\bibinfo{year}{2011})
  \bibinfo{pages}{1--12}.
%Type = Article
\bibitem[{Dodds and Watts(2005)}]{Dodds05}
\bibinfo{author}{P.~S. Dodds}, \bibinfo{author}{D.~J. Watts},
\newblock \bibinfo{title}{A generalized model of social and biological
  contagion},
\newblock \bibinfo{journal}{J. Theor. Biol.} \bibinfo{volume}{232}
  (\bibinfo{year}{2005}) \bibinfo{pages}{587--604}.
%Type = Article
\bibitem[{Centola et~al.(2007)Centola, Eguiluz, and Macy}]{Centola07a}
\bibinfo{author}{D.~Centola}, \bibinfo{author}{V.~M. Eguiluz},
  \bibinfo{author}{M.~W. Macy},
\newblock \bibinfo{title}{Cascade dynamics of complex propagation},
\newblock \bibinfo{journal}{Physica A} \bibinfo{volume}{374}
  (\bibinfo{year}{2007}) \bibinfo{pages}{449--456}.
%Type = Article
\bibitem[{Lopez-Pintado and Watts(2008)}]{LopezPintado08}
\bibinfo{author}{D.~Lopez-Pintado}, \bibinfo{author}{D.~J. Watts},
\newblock \bibinfo{title}{Social influence, binary decisions and collective
  dynamics},
\newblock \bibinfo{journal}{Rationality and Society} \bibinfo{volume}{20}
  (\bibinfo{year}{2008}) \bibinfo{pages}{399--443}.
%Type = Article
\bibitem[{Granovetter(1978)}]{Granovetter78}
\bibinfo{author}{M.~Granovetter},
\newblock \bibinfo{title}{Threshold models of collective behavior},
\newblock \bibinfo{journal}{Am. J. Sociol.} \bibinfo{volume}{83}
  (\bibinfo{year}{1978}) \bibinfo{pages}{1420--1443}.
%Type = Article
\bibitem[{Fowler and Christakis(2010)}]{Fowler10}
\bibinfo{author}{J.~H. Fowler}, \bibinfo{author}{N.~A. Christakis},
\newblock \bibinfo{title}{Cooperative behavior cascades in human social
  networks},
\newblock \bibinfo{journal}{Proc. Natl. Acad. Sci. U.S.A.}
  \bibinfo{volume}{107} (\bibinfo{year}{2010}) \bibinfo{pages}{5334--5338}.
%Type = Article
\bibitem[{Baldassarri and Bearmann(2007)}]{Baldassarri07}
\bibinfo{author}{D.~Baldassarri}, \bibinfo{author}{P.~Bearmann},
\newblock \bibinfo{title}{Dynamics of political polarization},
\newblock \bibinfo{journal}{Am. Sociol. Rev.} \bibinfo{volume}{72}
  (\bibinfo{year}{2007}) \bibinfo{pages}{784--811}.
%Type = Article
\bibitem[{Friedkin and Johnsen(1999)}]{Friedkin99}
\bibinfo{author}{N.~E. Friedkin}, \bibinfo{author}{E.~C. Johnsen},
\newblock \bibinfo{title}{Social influence networks and opinion change},
\newblock \bibinfo{journal}{Advances in Group Processes} \bibinfo{volume}{16}
  (\bibinfo{year}{1999}) \bibinfo{pages}{1--29}.
%Type = Article
\bibitem[{Hegselmann and Krause(2002)}]{Hegselmann02}
\bibinfo{author}{R.~Hegselmann}, \bibinfo{author}{U.~Krause},
\newblock \bibinfo{title}{Opinion dynamics and bounded confidence models,
  analysis, and simulation},
\newblock \bibinfo{journal}{Journal of Artificial Societies and Social
  Simulation} \bibinfo{volume}{5} (\bibinfo{year}{2002}).
%Type = Article
\bibitem[{Ben-Naim et~al.(2003)Ben-Naim, Krapivsky, Vazquez, and
  Redner}]{Ben-Naim03}
\bibinfo{author}{E.~Ben-Naim}, \bibinfo{author}{P.~L. Krapivsky},
  \bibinfo{author}{F.~Vazquez}, \bibinfo{author}{S.~Redner},
\newblock \bibinfo{title}{Unity and discord in opinion dynamics},
\newblock \bibinfo{journal}{Physica A} \bibinfo{volume}{330}
  (\bibinfo{year}{2003}) \bibinfo{pages}{99--106}.
%Type = Article
\bibitem[{Holme and Newman(2006)}]{Holme06}
\bibinfo{author}{P.~Holme}, \bibinfo{author}{M.~E.~J. Newman},
\newblock \bibinfo{title}{Nonequilibrium phase transition in the coevolution of
  networks and opinions},
\newblock \bibinfo{journal}{Phys. Rev. E} \bibinfo{volume}{74}
  (\bibinfo{year}{2006}) \bibinfo{pages}{1--5}.
%Type = Article
\bibitem[{Gil and Zanette(2006)}]{Gil06}
\bibinfo{author}{S.~Gil}, \bibinfo{author}{D.~H. Zanette},
\newblock \bibinfo{title}{Coevolution of agents and networks: Opinion spreading
  and community disconnection},
\newblock \bibinfo{journal}{Physics Letters A} \bibinfo{volume}{356}
  (\bibinfo{year}{2006}) \bibinfo{pages}{89--94}.
%Type = Article
\bibitem[{Kozma and Barrat(2008)}]{Kozma08}
\bibinfo{author}{B.~Kozma}, \bibinfo{author}{A.~Barrat},
\newblock \bibinfo{title}{Consensus formation on adaptive networks},
\newblock \bibinfo{journal}{Phys. Rev. E} \bibinfo{volume}{77}
  (\bibinfo{year}{2008}) \bibinfo{pages}{1--10}.
%Type = Article
\bibitem[{Mobilia(2011)}]{Mobilia11}
\bibinfo{author}{M.~Mobilia},
\newblock \bibinfo{title}{Fixation and polarization in a three-species opinion
  dynamics model},
\newblock \bibinfo{journal}{Europhys. Lett.} \bibinfo{volume}{95}
  (\bibinfo{year}{2011}) \bibinfo{pages}{1--6}.
%Type = Article
\bibitem[{McCullen et~al.(2011)McCullen, Ivanchenko, Shalfeev, and
  Gale}]{McCullen11}
\bibinfo{author}{N.~J. McCullen}, \bibinfo{author}{M.~V. Ivanchenko},
  \bibinfo{author}{V.~D. Shalfeev}, \bibinfo{author}{W.~F. Gale},
\newblock \bibinfo{title}{A dynamical model of decision-making behaviour in a
  network of consumers with applications to energy choices},
\newblock \bibinfo{journal}{Int. J. Bif. Chaos} \bibinfo{volume}{21}
  (\bibinfo{year}{2011}) \bibinfo{pages}{2467--2480}.
%Type = Article
\bibitem[{Durrett et~al.(2011)Durrett, Gleeson, Lloyd, Mucha, Shi, Sivakoff,
  Socolar, and Varghese}]{Durrett11}
\bibinfo{author}{R.~Durrett}, \bibinfo{author}{J.~P. Gleeson},
  \bibinfo{author}{A.~L. Lloyd}, \bibinfo{author}{P.~J. Mucha},
  \bibinfo{author}{F.~Shi}, \bibinfo{author}{D.~Sivakoff},
  \bibinfo{author}{J.~E.~S. Socolar}, \bibinfo{author}{C.~Varghese},
\newblock \bibinfo{title}{Graph fission in an evolving voter model},
\newblock \bibinfo{journal}{Proc. Natl. Acad. Sci. U.S.A.}
  \bibinfo{volume}{109} (\bibinfo{year}{2011}) \bibinfo{pages}{3683--3687}.
%Type = Article
\bibitem[{Shoham and Tennenholtz(1997)}]{Shoham97}
\bibinfo{author}{Y.~Shoham}, \bibinfo{author}{M.~Tennenholtz},
\newblock \bibinfo{title}{On the emergence of social conventions: modeling,
  analysis, and simulations},
\newblock \bibinfo{journal}{Artificial Intelligence} \bibinfo{volume}{94}
  (\bibinfo{year}{1997}) \bibinfo{pages}{139--166}.
%Type = Article
\bibitem[{Friedkin(2001)}]{Friedkin01}
\bibinfo{author}{N.~E. Friedkin},
\newblock \bibinfo{title}{Norm formation in social influence networks},
\newblock \bibinfo{journal}{Soc. Networks} \bibinfo{volume}{23}
  (\bibinfo{year}{2001}) \bibinfo{pages}{167--189}.
%Type = Article
\bibitem[{Centola et~al.(2005)Centola, Willer, and M.}]{Centola05}
\bibinfo{author}{D.~Centola}, \bibinfo{author}{R.~Willer},
  \bibinfo{author}{M.~M.},
\newblock \bibinfo{title}{The emperor’s dilemma: A computational model of
  self-enforcing norms},
\newblock \bibinfo{journal}{Am. J. Sociol.} \bibinfo{volume}{110}
  (\bibinfo{year}{2005}) \bibinfo{pages}{1009--1040}.
%Type = Article
\bibitem[{Castellano et~al.(2009)Castellano, Fortunato, and
  Loreto}]{Castellano09}
\bibinfo{author}{C.~Castellano}, \bibinfo{author}{S.~Fortunato},
  \bibinfo{author}{V.~Loreto},
\newblock \bibinfo{title}{Statistical physics of social dynamics},
\newblock \bibinfo{journal}{Rev. Mod. Phys.} \bibinfo{volume}{81}
  (\bibinfo{year}{2009}) \bibinfo{pages}{591--646}.
%Type = Article
\bibitem[{McPherson et~al.(2001)McPherson, Smith-Lovin, and Cook}]{McPherson01}
\bibinfo{author}{M.~McPherson}, \bibinfo{author}{L.~Smith-Lovin},
  \bibinfo{author}{J.~M. Cook},
\newblock \bibinfo{title}{Birds of a feather: Homophily in social networks},
\newblock \bibinfo{journal}{Ann. Rev. Soc.} \bibinfo{volume}{27}
  (\bibinfo{year}{2001}) \bibinfo{pages}{415--444}.
%Type = Book
\bibitem[{Lazarsfeld and Merton(1954)}]{Lazarsfeld54}
\bibinfo{author}{P.~F. Lazarsfeld}, \bibinfo{author}{R.~K. Merton},
  \bibinfo{title}{Friendship as a social process: a substantive and
  methodological analysis}, \bibinfo{publisher}{Octagon Books},
  \bibinfo{address}{New York, USA}, \bibinfo{year}{1954}.
%Type = Article
\bibitem[{Centola et~al.(2007)Centola, Gonz\'{a}lez-Avella, Egu\'{i}luz, and
  San~Miguel}]{Centola07}
\bibinfo{author}{D.~Centola}, \bibinfo{author}{J.~C. Gonz\'{a}lez-Avella},
  \bibinfo{author}{V.~M. Egu\'{i}luz}, \bibinfo{author}{M.~San~Miguel},
\newblock \bibinfo{title}{Homophily, cultural drift, and the co-evolution of
  cultural groups},
\newblock \bibinfo{journal}{J. Conflict Resolut.} \bibinfo{volume}{51}
  (\bibinfo{year}{2007}) \bibinfo{pages}{905--929}.
%Type = Article
\bibitem[{Gross and Blasius(2007)}]{Gross07}
\bibinfo{author}{T.~Gross}, \bibinfo{author}{B.~Blasius},
\newblock \bibinfo{title}{Adaptive coevolutionary networks: a review},
\newblock \bibinfo{journal}{J. R. Soc. Interface} \bibinfo{volume}{5}
  (\bibinfo{year}{2007}) \bibinfo{pages}{259--271}.
%Type = Book
\bibitem[{Sherif(1936)}]{Sherif36}
\bibinfo{author}{M.~Sherif}, \bibinfo{title}{The Psychology of Social Norms},
  \bibinfo{publisher}{Harper}, \bibinfo{address}{Oxford, England},
  \bibinfo{year}{1936}.
%Type = Incollection
\bibitem[{Campbell(1963)}]{Campbell61}
\bibinfo{author}{D.~T. Campbell},
\newblock \bibinfo{title}{Conformity in psychology's theories of acquired
  behavioural disposition},
\newblock in: \bibinfo{booktitle}{Conformity and Deviation},
  \bibinfo{publisher}{Harper}, \bibinfo{address}{New York},
  \bibinfo{year}{1963}.
%Type = Book
\bibitem[{Collins and Ashmore(1970)}]{Collins70}
\bibinfo{author}{B.~E. Collins}, \bibinfo{author}{R.~D. Ashmore},
  \bibinfo{title}{Social Psychology: Social Influence, attitude change, group
  processes and prejudice}, \bibinfo{publisher}{Addison-Wesley Publishing
  Company}, \bibinfo{address}{Reading, Massachusetts}, \bibinfo{year}{1970}.
%Type = Incollection
\bibitem[{Asch(1951)}]{Asch51}
\bibinfo{author}{S.~E. Asch},
\newblock \bibinfo{title}{Effects of group pressure upon the modification and
  distortion of judgment},
\newblock in: \bibinfo{booktitle}{Groups, Leadership and Men},
  \bibinfo{publisher}{The Carnegie Press}, \bibinfo{address}{Pittsburgh, USA},
  \bibinfo{year}{1951}.
%Type = Article
\bibitem[{Milgram(1965)}]{Milgram65}
\bibinfo{author}{S.~Milgram},
\newblock \bibinfo{title}{Liberating effects of group pressure},
\newblock \bibinfo{journal}{J. Pers. Soc. Psychol.} \bibinfo{volume}{1}
  (\bibinfo{year}{1965}) \bibinfo{pages}{127--134}.
%Type = Article
\bibitem[{Flache and Macy(2011)}]{Flache11}
\bibinfo{author}{A.~Flache}, \bibinfo{author}{M.~W. Macy},
\newblock \bibinfo{title}{Small worlds and cultural polarization},
\newblock \bibinfo{journal}{J. Math. Soc.} \bibinfo{volume}{35}
  (\bibinfo{year}{2011}) \bibinfo{pages}{146--176}.
%Type = Article
\bibitem[{Aral et~al.(2009)Aral, Muchnik, and Sundararajan}]{Aral09}
\bibinfo{author}{S.~Aral}, \bibinfo{author}{L.~Muchnik},
  \bibinfo{author}{A.~Sundararajan},
\newblock \bibinfo{title}{Distinguishing influence-based contagion from
  homophily-driven diffusion in dynamic networks},
\newblock \bibinfo{journal}{Proc. Natl. Acad. Sci. U.S.A.}
  \bibinfo{volume}{106} (\bibinfo{year}{2009}) \bibinfo{pages}{21544--21549}.
%Type = Article
\bibitem[{Shalizi and Thomas(2011)}]{Shalizi11}
\bibinfo{author}{C.~R. Shalizi}, \bibinfo{author}{A.~C. Thomas},
\newblock \bibinfo{title}{Homophily and contagion are generically confounded in
  observational social network studies},
\newblock \bibinfo{journal}{Sociological Methods \& Research}
  \bibinfo{volume}{40} (\bibinfo{year}{2011}) \bibinfo{pages}{211--239}.
%Type = Article
\bibitem[{Carley(1991)}]{Carley91}
\bibinfo{author}{K.~Carley},
\newblock \bibinfo{title}{A theory of group stability},
\newblock \bibinfo{journal}{Am. Sociol. Rev.} \bibinfo{volume}{56}
  (\bibinfo{year}{1991}) \bibinfo{pages}{331--354}.
%Type = Article
\bibitem[{Mark(1998)}]{Mark98}
\bibinfo{author}{N.~Mark},
\newblock \bibinfo{title}{Beyond individual differences: Social differentiation
  from first principles},
\newblock \bibinfo{journal}{Am. Sociol. Rev.} \bibinfo{volume}{63}
  (\bibinfo{year}{1998}) \bibinfo{pages}{309--330}.
%Type = Article
\bibitem[{Axelrod(1997)}]{Axelrod97}
\bibinfo{author}{R.~Axelrod},
\newblock \bibinfo{title}{The dissemination of culture},
\newblock \bibinfo{journal}{J. Conflict Resolut.} \bibinfo{volume}{41}
  (\bibinfo{year}{1997}) \bibinfo{pages}{203--226}.
%Type = Inproceedings
\bibitem[{Macy et~al.(2003)Macy, Kitts, Flache, and Benard}]{Macy03}
\bibinfo{author}{M.~W. Macy}, \bibinfo{author}{J.~A. Kitts},
  \bibinfo{author}{A.~Flache}, \bibinfo{author}{S.~Benard},
\newblock \bibinfo{title}{Polarization in dynamic networks: A {H}opfield model
  of emergent structure},
\newblock in: \bibinfo{booktitle}{Dynamic Social Network Modeling and
  Analysis}, \bibinfo{publisher}{The National Acadamy Press},
  \bibinfo{address}{Washington, D.C.}, \bibinfo{year}{2003}.
%Type = Article
\bibitem[{Klemm et~al.(2005)Klemm, M., Toral, and San~Miguel}]{Klemm05}
\bibinfo{author}{K.~Klemm}, \bibinfo{author}{E.~V. M.},
  \bibinfo{author}{R.~Toral}, \bibinfo{author}{M.~San~Miguel},
\newblock \bibinfo{title}{Globalization, polarization and cultural drift},
\newblock \bibinfo{journal}{J Econ. Dyn. Control} \bibinfo{volume}{29}
  (\bibinfo{year}{2005}) \bibinfo{pages}{321--334}.
%Type = Article
\bibitem[{Morgane et~al.(2005)Morgane, Galler, and Mokler}]{Morgane05}
\bibinfo{author}{P.~J. Morgane}, \bibinfo{author}{J.~R. Galler},
  \bibinfo{author}{D.~J. Mokler},
\newblock \bibinfo{title}{A review of systems and networks of the limbic
  forebrain/limbic midbrain},
\newblock \bibinfo{journal}{Progress in Neurobiology} \bibinfo{volume}{75}
  (\bibinfo{year}{2005}) \bibinfo{pages}{143--160}.
%Type = Article
\bibitem[{Mathews and Gilliland(1999)}]{Mathews99}
\bibinfo{author}{G.~Mathews}, \bibinfo{author}{K.~Gilliland},
\newblock \bibinfo{title}{The personality theories of {H}. {J}. {E}ysenck and
  {J}. {A}. {G}ray: a comparative review},
\newblock \bibinfo{journal}{Pers. Indiv. Differ.} \bibinfo{volume}{26}
  (\bibinfo{year}{1999}) \bibinfo{pages}{583--626}.
%Type = Book
\bibitem[{Eysenck(1967)}]{Eysenck67}
\bibinfo{author}{H.~J. Eysenck}, \bibinfo{title}{The biological basis of
  personality}, \bibinfo{publisher}{Thomas}, \bibinfo{address}{Springfield},
  \bibinfo{year}{1967}.
%Type = Book
\bibitem[{Gray(1987)}]{Gray87}
\bibinfo{author}{J.~A. Gray}, \bibinfo{title}{The psychology of fear and
  stress}, \bibinfo{publisher}{Cambridge University Press},
  \bibinfo{address}{Cambridge}, \bibinfo{edition}{2nd} edition,
  \bibinfo{year}{1987}.
%Type = Article
\bibitem[{Nakao and Mihailov(2010)}]{nakao10}
\bibinfo{author}{H.~Nakao}, \bibinfo{author}{A.~S. Mihailov},
\newblock \bibinfo{title}{Turing patterns in network-organized
  activator–inhibitor systems},
\newblock \bibinfo{journal}{Nat. Phys.} \bibinfo{volume}{6}
  (\bibinfo{year}{2010}) \bibinfo{pages}{544--550}.
%Type = Unpublished
\bibitem[{Grindrod and Parsons(2012)}]{Parsons12}
\bibinfo{author}{P.~Grindrod}, \bibinfo{author}{M.~Parsons},
  \bibinfo{title}{Complex dynamics in a model of social norms},
  \bibinfo{year}{2012}.
%Type = Article
\bibitem[{DeLellis et~al.(2010)DeLellis, diBernardo, Garofalo, and
  Porfiri}]{DeLellis10}
\bibinfo{author}{P.~DeLellis}, \bibinfo{author}{M.~diBernardo},
  \bibinfo{author}{M.~Garofalo}, \bibinfo{author}{M.~Porfiri},
\newblock \bibinfo{title}{Evolution of complex networks via edge snapping},
\newblock \bibinfo{journal}{IEEE Transactions on Circuits and Systems}
  \bibinfo{volume}{57} (\bibinfo{year}{2010}) \bibinfo{pages}{2132--2143}.

\end{thebibliography}






% \end{thebibliography}

\end{document}